\documentclass[superscriptaddress]{revtex4-2}
\RequirePackage{jabbrv}
\usepackage[english]{babel}

\usepackage{graphicx}
\DeclareGraphicsExtensions{{.pdf}}
\usepackage{hyperref}

\begin{document}

\title{Fabrication of high-$Q$ defect-free optical nanofiber photonic crystal resonators}
\author{Tomofumi Tanaka}
\author{Takahiro Suzuki}
\author{Owen Mao}
\author{Samuel K. Ruddell}
\author{Karen E. Webb}
\affiliation{Department of Applied Physics, Waseda University, 3-4-1 Okubo, Shinjuku, Tokyo 169-8555, Japan}
\author{Takao Aoki}
\email{takao@waseda.jp}
\affiliation{Department of Applied Physics, Waseda University, 3-4-1 Okubo, Shinjuku, Tokyo 169-8555, Japan}
\affiliation{RIKEN Center for Quantum Computing (RQC), Wako, Saitama 351-0198, Japan}

\begin{abstract}
We demonstrate the fabrication of defect-free optical-nanofiber photonic-crystal Fabry-P\'erot resonators with quality factors exceeding $10^7$ using single-shot femtosecond laser ablation. An investigation of the nonlinear optical properties reveals that thermo-optic effects dominate within the entire cavity bandwidth, even when interrogating with pulses one order of magnitude shorter than the 6.6~$\mu$s thermal cutoff time. The combination of high-$Q$ and small mode volume of these resonators could facilitate the creation of high-speed quantum nodes for cavity QED based quantum computing and networking, as well as low-power in-line fiber optical switches.
\end{abstract}

\maketitle

\section{Introduction}~\label{sec:introduction}

\noindent Optical microresonators enable the creation of low-power, high-speed, compact photonic devices~\cite{vahala2003}. Their ability to strongly confine light, both spatially and temporally, allows for high intracavity intensities with only moderate input power, which can consequently enable strong light-matter interactions. This can provide major advantages for applications such as optical filtering and switching~\cite{pollinger2009, pollinger2010}, lasing~\cite{kato2022}, sensing~\cite{zhang2015}, nonlinear optics~\cite{kippenberg2011}, and cavity quantum electrodynamics (QED)~\cite{kato2015, ruddell2017, nayak2018, kato2019, ruddell2020}. 

Nanofiber-based optical microresonators have recently emerged as a particular platform of interest for facilitating cavity QED based quantum computing and networking, as well as for nonlinear optics~\cite{wuttke2012}. The demand for higher quality factors~($Q$) has necessarily driven vast improvements in fabrication techniques~\cite{ruddell2020, kato2022, horikawa2025}, such that the nonlinear properties of the resulting resonators can begin to approach that of some whispering gallery mode (WGM) type microresonators. The in-line fiber nature of these nanofiber-based resonators also precludes issues with mechanical stability and alignment.

Various implementations of optical nanofiber-based resonators have been demonstrated, including fiber Fabry-P\'erot cavities~\cite{kato2015, kato2019}, fiber ring cavities~\cite{jones2016, schneeweiss2017, ruddell2017}, as well as photonic crystal~(PhC) cavities integrated into the nanofiber itself, fabricated using focussed ion beam milling~\cite{nayak2011, schell2015, li2017, takashima2019, tashima2022, romagnoli2020}, femtosecond laser ablation~\cite{nayak2013, nayak2014, keloth2017, nayak2017, wang2019b}, and those implemented through the use of external gratings~\cite{sadgrove2013, yalla2014, yalla2020, yalla2022}.

In this paper, we report on the fabrication of a defect-free PhC Fabry-P\'erot resonator on an optical nanofiber with a $Q$-factor exceeding $10^7$ using single-shot femtosecond laser ablation. We investigate the nonlinear properties of these resonators, and find that the thermo-optic effect dominates over the entire cavity bandwidth, even when using pulses one order of magnitude shorter than the 6.6~$\mu$s thermal cutoff time. The combination of high-$Q$ and small mode volume of these devices could facilitate the creation of high-speed cavity QED-based quantum nodes. Additionally, due to their strong thermo-optic response, these resonators could be used as low-power in-line fiber optical switches.

\section{Nanofiber PhC Resonator Fabrication}~\label{sec:fabrication}
\noindent To fabricate a defect-free optical-nanofiber PhC cavity, we initially taper standard single-mode optical fiber (SM800, Fibercore) to form a nanofiber with a diameter of 500~nm and waist length of 13~mm using the flame-brush technique~\cite{birks1992, nagai2014, ruddell2020}. To increase mechanical stability, the shape of the tapered region is optimized to minimize the overall taper length while ensuring that the taper transition remains adiabatic~\cite{nagai2014}. During the tapering process, we slowly lower the flow rate of $\textrm{H}_2$ and $\textrm{O}_2$ to prevent bending due to pressure from the torch. The taper transmission is monitored using an intensity stabilised 852~nm laser, and the waist diameter is monitored by observing the spectrogram of a 688-nm laser during the pulling process~\cite{ruddell2020}. Once taper fabrication is complete, the tapered fiber is slightly tensioned and fixed to a jig by UV~curing glue in preparation for nanofiber PhC cavity fabrication.

\begin{figure}[htb]
\begin{center}
\includegraphics{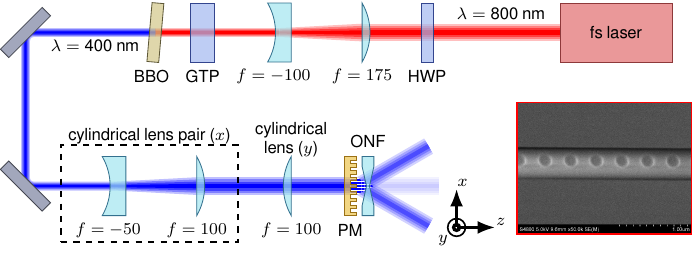}
\end{center}
\vspace{-10pt}
\caption{Experimental setup for fabricating PhC cavities on nanofibers by femtosecond laser ablation. HWP: half-wave plate, GTP: Glan-Taylor polarizer, BBO: barium borate crystal, PM: phase mask, ONF: optical nanofiber. Lens focal length units are given in mm. Inset: SEM image of a section of a PhC cavity fabricated using this setup.}
\label{fig:exposure_system}
\end{figure}

The nanofiber PhC cavity fabrication setup is shown in Fig.~\ref{fig:exposure_system}. A femtosecond laser (Coherent Astrella-F-1K) capable of producing $<100$~fs pulses with a repetition rate of 1~kHz at a central wavelength of 800~nm is frequency doubled to 400~nm by a barium borate (BBO) crystal. The output beam is initially expanded in the $x$-direction by a pair of cylindrical lenses, then focussed to a line using a cylindrical lens in the $y$-direction to maximise the intensity of light along the nanofiber region. The beam passes through a 30~mm wide phase mask, optimized to diffract light into both the $\pm$1 orders, where the interference of these two orders forms an interference pattern to be imaged on to the nanofiber. The 372.5~nm period of the interference pattern is designed to create nanofiber Bragg gratings with a stop band near 852~nm for 500~nm diameter nanofibers.

As the nanofiber must be precisely aligned with respect to the phase mask prior to exposure, we employ a 5-axis micrometer stage for positioning, allowing for translation and rotation of the nanofiber in all relevant dimensions. During alignment, the fiber is observed from above to confirm alignment and position relative to the phase mask. Additionally, the laser is configured to output low-power light, with each of the $\pm1$~diffraction orders being observed from behind the nanofiber using a camera and a photodetector. This enables us to ensure that light is diffracted symmetrically by the nanofiber, and that the nanofiber is aligned to the position of peak power.

Once the nanofiber has been aligned, the beam is blocked using a mechanical shutter. The laser is then switched to single-shot mode, where single femtosecond pulses can be triggered. The mechanical shutter is then opened, and the nanofiber is exposed to a single $\sim$530~mW femtosecond~laser pulse. The circular cross section of the nanofiber causes the front face to act as a lens, focussing light onto the rear side of the nanofiber, where material is ablated. Due to the intensity diffraction pattern produced by the phase mask, a periodic series of craters are formed, resulting in a nanofiber PhC Bragg grating (see inset of Fig.~\ref{fig:exposure_system}). The size of the individual craters smoothly varies along the nanofiber due to the Gaussian intensity distribution of the laser beam profile, and hence the effective refractive index of the nanofiber guided mode also varies. As the distribution of crater sizes is symmetric along the nanofiber, a defect-free cavity can be formed between two regions of grating with matching effective refractive index. This type of defect-free nanofiber cavity potentially has several advantages over other types of nanofiber-based cavities. The use of single-shot femtosecond laser ablation allows the fabrication process to be essentially impervious to mechanical vibrations, and the defect-free nature of the cavity ensures that there is no large jump in effective refractive index at a defect boundary, which could induce loss.

An example of the transmission spectrum of a fabricated nanofiber PhC cavity is shown in Fig.~\ref{fig:spm_setup}(b), consisting of a reflection band spanning a few nanometers and containing narrow cavity resonances. Due to the tight confinement of light within the small nanofiber PhC cavity mode volume, as well as the high-$Q$ of the resonances, we expect these resonators to display a high degree of nonlinearity~\cite{pollinger2009, pollinger2010}. We therefore perform an investigation of the nonlinear properties of these cavities through self-phase modulation and cross-phase modulation measurements.

\section{Characterization of Nanofiber PhC Resonators}~\label{sec:characterization}
\begin{figure}[htb]
\begin{center}
\includegraphics{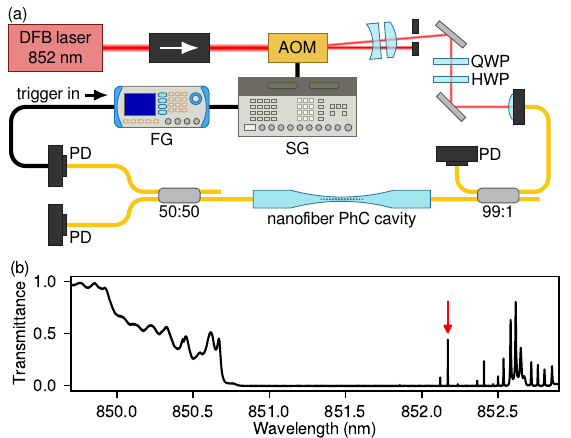}
\end{center}
\vspace{-10pt}
\caption{(a) Experimental setup used for measuring SPM in a nanofiber PhC cavity. AOM: acousto-optic modulator, QWP: quarter-wave plate, HWP: half-wave plate, PD: photodetector, FG: function generator, SG: signal generator. (b) Transmission spectrum of the PhC cavity used for SPM measurements. Red arrow indicates the target resonance.}
\label{fig:spm_setup}
\end{figure}

\begin{figure}[htb]
\begin{center}
\includegraphics{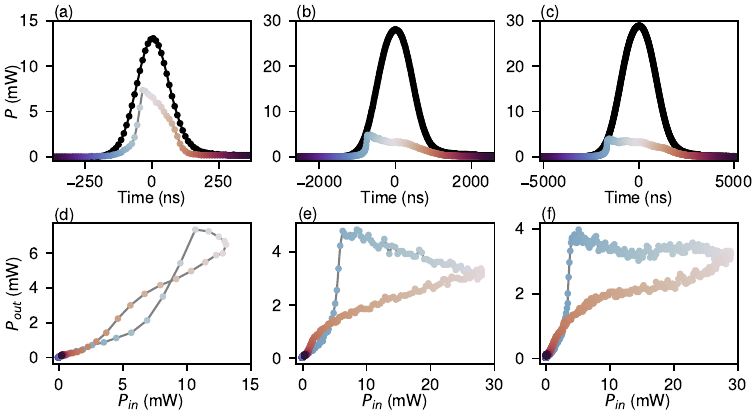}
\end{center}
\vspace{-10pt}
\caption{Results of the SPM experiment shown in Fig. 2. (a)-(c) Input pulse (black dots) and corresponding cavity output (colored dots). The input pulse widths and detunings are (a)~147~ns and $\delta=1.78$, (b)~1.02~$\mu$s and $\delta=1.6$, and (c)~2.02~$\mu$s and $\delta=1.71$. (d-f) Cavity output versus cavity input corresponding to (a-c).}
\label{fig:spm_result}
\end{figure}

\noindent For resonator materials exhibiting a third-order susceptibility $\chi^{(3)}$, the refractive index is intensity dependent due to the optical Kerr effect, given by \mbox{$n = n_0 + n_2\, I$}, where $n_0$ and $n_2$ are the linear and second-order nonlinear refractive indices, respectively, and $I$ is the intensity of light. By injecting a short pulse of light into a cavity resonance, the resulting change in refractive index leads to a shift of the resonance frequency, and above a certain threshold power, to optical bistability. 
This threshold power is proportional to $V_\mathrm{mode}/Q^2$, where $V_\mathrm{mode}$ is the resonator mode volume~\cite{joannopoulos2008}.

For cavities with a high quality factor, thermo-optic effects can also lead to a similar bistability phenomenon, where the temperature dependence of the resonance wavelength $\lambda_0$ is given by
\begin{equation}
\frac{\mathrm{d}\lambda}{\lambda_0} = \left(\alpha + \frac{1}{n_0}\frac{\mathrm{d}n}{\mathrm{d}T}\right)\mathrm{d}T,
\label{eq:photothermal}
\end{equation}
where $\alpha$ is the thermal expansion coefficient and $(1/n_0)/(\mathrm{d}n/\mathrm{d}T)$ is the thermo-optic coefficient~\cite{hutner2020}. Typically, the photothermal effect is characterized by a much slower response time than that of the Kerr effect.

We utilize a nanofiber PhC cavity with the transmission spectrum shown in Fig.~\ref{fig:spm_setup}(b), and characterize the target resonance indicated by the red arrow, having a FWHM of $\Delta\nu=32$~MHz at 852.35~nm, the wavelength of the cesium D$_2$ transition. The corresponding 
quality factor is \mbox{$Q = 1.1\times10^7$}, with an intrinsic quality factor of \mbox{$Q_\mathrm{i}=2.9\times10^7$}.

We investigate the nonlinear properties of this resonator using a self-phase modulation (SPM) technique, using the setup shown in Fig.~\ref{fig:spm_setup}(a). Here, an 852~nm DFB laser is slowly scanned from lower frequency into the target cavity resonance at low power. When the shoulder of the resonance is reached, we immediately trigger a rapid Gaussian intensity modulation using an AOM. The time evolution of both the input and output power is monitored, with results shown in Fig.~\ref{fig:spm_result}. We observed a nonlinear response of $P_\mathrm{out}$ for all input pulse widths, ranging from 147~ns up to 2~$\mu$s. Furthermore, bifurcation of the hysteresis loop was observed, indicating bistable behaviour. 

Contrary to the theoretical model for optical bistability based purely on the Kerr effect, the observed bistable curves did not merge after branching, and instead a new clockwise branch occured at large $P_\mathrm{in}$, as shown in Fig.~\ref{fig:spm_result}(d--f). When viewed as a function of time, $P_\mathrm{out}$ exhibited a larger than expected rise during the first half of the pulse, referred to as overshoot. This overshoot was more pronounced for longer pulse widths, indicating that it may instead be due to the thermo-optic effect. Due to the relatively slow response time, we expect that the thermo-optic effect will weaken with shorter pulse widths, however we observed that the overshoot did not completely disappear, even when using pulses as short as 147~ns, the shortest we were capable of generating using this setup. This implies that the thermo-optic effect is dominant even on such short time scales.

At longer pulse-widths, where the thermo-optic effect is strongest, the threshold for optical bistability is reduced compared to that of the pure Kerr effect. 
Reducing the peak input power of the pulse, either directly or by increasing the frequency detuning $\delta$ from resonance, reduces the overall nonlinear response, and does not help to resolve the threshold due solely to the Kerr effect. At large detunings, the nonlinearity is not observed at all, and the hysteresis curve becomes linear. 

\begin{figure}[htb]
\begin{center}
\includegraphics{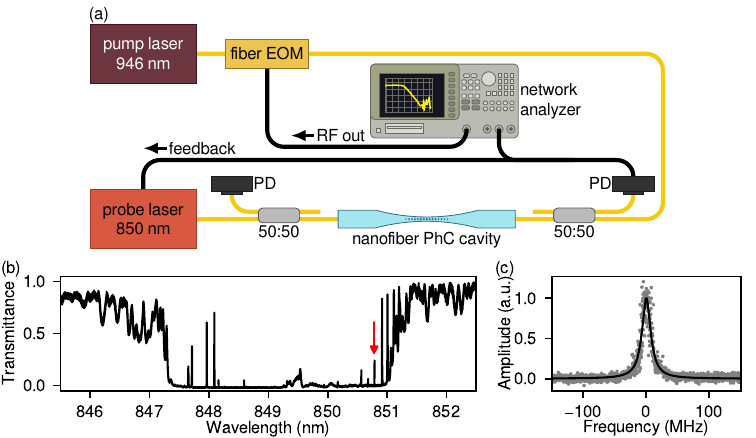}
\end{center}
\vspace{-10pt}
\caption{(a) Experimental setup used for measuring XPM in nanofiber PhC cavities. EOM: electro-optic modulator, PD: photodetector. (b) Transmission spectrum of the PhC cavity used in XPM measurements. Red arrow indicates the target resonance to be investigated. (c) Linewidth measurement (gray dots) of the target resonance in (b). Black line is the fitted Lorentzian linewidth, with a FWHM of 16.6~MHz.}
\label{fig:xpm_setup}
\end{figure}

\begin{figure}[htb]
\begin{center}
\includegraphics{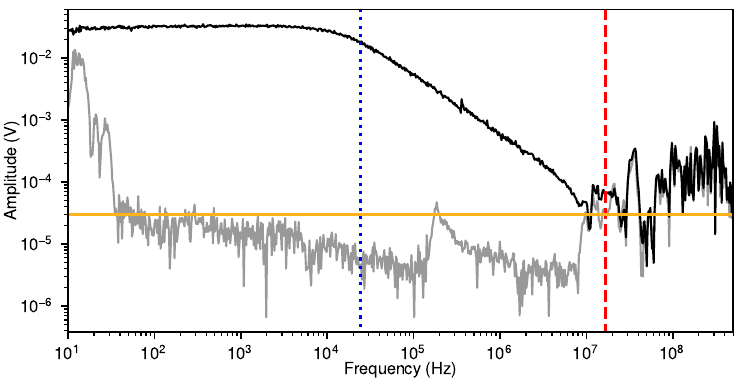}
\end{center}
\vspace{-10pt}
\caption{Result of the XPM measurement. Black line is the measured frequency response with both the probe and pump lasers, while gray line is the measured frequency response of only the pump laser. Orange horizontal line corresponds to the expected Kerr effect plateau, blue dotted line corresponds to the cutoff frequency of the photothermal effect at 24~kHz, and red dashed line corresponds to the cavity linewidth of 16.6~MHz.}
\label{fig:xpm_result}
\end{figure}

Observing the response due solely to the Kerr effect using SPM techniques would therefore require the use of extremely short pulse widths such that thermal effects are eliminated, with the caveat that the pulse width lies within the bandwidth of the cavity resonance. We therefore study the frequency dependence of the nonlinear response using cross-phase modulation (XPM) to determine the cutoff frequency of the thermo-optic effect. For XPM, light with two differing wavelengths, a pump and a probe, are incident onto a medium. By modulating the pump light intensity, the refractive index perceived by the probe light is also modulated, resulting in phase modulation of the probe light. 

The apparatus used for the XPM measurement is shown in Fig.~\ref{fig:xpm_setup}(a). We utilize a $\sim$3.5~mW pump with a wavelength of 946~nm, which falls outside the stopband of the PhC resonator, and additionally avoids noise due to unwanted nonlinear processes within the region of interest, such as stimulated Raman scattering. Low power probe light is tuned to the target resonance, and the laser frequency is locked to the shoulder of the resonance peak using a relatively low bandwidth feedback loop. This allows us to modulate the intensity of the pump laser and observe the response in intensity of the probe light to determine the frequency dependence of nonlinear effects within the cavity.

At low frequencies, we expect the refractive index change due to XPM to be a combination of thermal and Kerr nonlinear processes. As the modulation frequency increases, the response due to the thermal effect will reduce beyond the thermal cutoff frequency. For typical microresonators, the thermo-optic response essentially disappears at high frequencies, and the refractive index change due to the Kerr effect can be directly measured~\cite{rokhsari2005, pollinger2010, gao2022}.

We utilize the nanofiber PhC cavity shown in Fig.~\ref{fig:xpm_setup}(b), and characterize the target resonance indicated by the red arrow having a FWHM of \mbox{$\Delta\nu=16.6$~MHz} at 850.8~nm, as shown in Fig.~\ref{fig:xpm_setup}(c), corresponding to a quality factor of \mbox{$Q = 2.1\times10^7$} and an intrinsic quality factor of \mbox{$Q_\mathrm{i} = 4.2\times10^7$}. The results of the XPM measurement are shown in Fig.~\ref{fig:xpm_result}, where the intensity of the pump laser has been modulated by $10\%$. We determine the cutoff frequency for the thermo-optic effect to be 24~kHz, corresponding to a characteristic time of 6.6~$\mu$s. We also observe that due to the high-$Q$ and small mode volume of the resonator, the thermo-optic effect dominates across the entire cavity bandwith. 

Although the response due to the Kerr effect cannot be measured directly for our system, we are able to obtain an estimate of the relative response level by calculating the expected change in refractive index due to the intensity modulation of the pump laser, and then considering the subsequent frequency shift of the resonance. For pure Kerr nonlinearity, the change in refractive index due to XPM is given by \mbox{$\Delta n_\mathrm{probe} = 2\, n_2\, I_\mathrm{pump}$}. The resulting estimate is plotted in Fig.~\ref{fig:xpm_result}, where we can observe that at low frequencies, the response due to the thermo-optic effect is approximately 1000 times larger than that of the Kerr effect. 

\section{Discussion}~\label{sec:discussion}
\noindent For a given in-coupled power, the intra-cavity intensity is proportional to $Q/V_{\mathrm{mode}}$, therefore producing resonators with small mode volumes can be desirable for applications requiring strong light-matter interactions. For the defect-free nanofiber PhC resonators presented here, the cross-sectional mode area can be established from measurements performed during nanofiber fabrication, however the effective cavity length depends on the precise structure of the PhC Bragg grating, and is therefore more difficult to quantitatively determine. The large free spectral range of these resonators, combined with the large dispersion of the PhC Bragg grating, prevents the cavity length from being determined by directly measuring the cavity spectrum. One alternative method involves measuring the power threshold for optical bistability due to the Kerr effect, however, this method is confounded by the large thermal response of these resonators, presumably due to their high $Q/V_{\mathrm{mode}}$ ratio combined with their large surface-to-volume ratio. Nevertheless, we expect the cavity length to be on the order of a few~mm based on the fabrication method and physical structure of the resonator, and therefore estimate the cavity mode volume to be on the order of $10^3~\mu \textrm{m}^3$.

In previous works, various methods have been utilized to fabricate nanofiber PhC cavities. While focussed ion milling techniques allow for fabrication of essentially arbitrary PhC structures, issues with contamination during fabrication limit the maximum achievable quality~\cite{romagnoli2020}. Femtosecond laser ablation has enabled optical production of nanofiber PhC Bragg gratings~\cite{nayak2013}, and subsequently the fabrication of both defect-free and defect-induced nanofiber PhC cavities~\cite{nayak2014, keloth2017}. For the defect-free nanofiber PhC cavities produced, resonances with quality factors of up to \mbox{$Q \approx 2\times10^6$} have been observed~\cite{nayak2014}, and therefore our cavities represent an order of magnitude increase in quality factor over previous works. 


\section{Conclusion}~\label{sec:conclusion}
\noindent We have fabricated multiple defect-free PhC nanofiber resonators containing resonances with quality factors exceeding $10^7$ by using femtosecond laser ablation. We performed an investigation of the nonlinear properties of these resonators by employing both SPM and XPM techniques. We determined the cutoff frequency of the thermo-optic response to be 24~kHz, and saw that these effects dominate across the entire bandwidth of the cavity. These resonators could potentially be used as a platform for creating high-speed cavity QED based quantum nodes, which could then further be integrated into a distributed quantum computing network. Additionally, these resonators could facilitate the fabrication of low-power, high-speed, in-line, fiber optical switches.


\end{document}